# Wideband Self-Adaptive RF Cancellation Circuit for Full-Duplex Radio: Operating Principle and Measurements

Timo Huusari*, Yang-Seok Choi†, Petteri Liikkanen*, Dani Korpi*, Shilpa Talwar‡, and Mikko Valkama*

*Department of Electronics and Communications Engineering, Tampere University of Technology, Finland
e-mail: timo.huusari@tut.fi, petteri.liikkanen@tut.fi, dani.korpi@tut.fi, mikko.e.valkama@tut.fi

†Intel Corporation, Hillsboro, Oregon, USA, e-mail: yang-seok.choi@intel.com.

‡Intel Corporation, Santa Clara, California, USA, e-mail: shilpa.talwar@intel.com.

***Abstract*—This paper presents a novel RF circuit architecture for self-interference cancellation in inband full-duplex radio transceivers . The developed canceller is able to provide wideband cancellation with waveform bandwidths in the order of 100 MHz or beyond and contains also self-adaptive or self-healing features enabling automatic tracking of time-varying self-interference channel characteristics. In addition to architecture and operating principle descriptions, we also provide actual RF measurements at 2.4 GHz ISM band demonstrating the achievable cancellation levels with different bandwidths and when operating in different antenna configurations and under low-cost highly nonlinear power amplifier. In a very challenging example with a 100 MHz waveform bandwidth, around 41 dB total cancellation is obtained while the corresponding cancellation figure is close to 60 dB with the more conventional 20 MHz carrier bandwidth. Also, efficient tracking in time-varying reflection scenarios is demonstrated.**

## I. Introduction

Inband full-duplex radio communications, where co-located transmitter and receiver are operating simultaneously at the same center-frequency, is currently under active research. This has the potential to offer various benefits over classical time-domain duplexing (TDD) and frequency domain duplexing (FDD) based solutions, such as increased spectral efficiency, real-time spectrum monitoring capabilities while transmitting and simplified radio network planning in terms of frequencies [1]–[4]. One of the key challenges on the implementation side, however, is the ability to cancel the powerful self-interference stemming from the transmit signal coupling to the receiver chain [1], [3]–[7]. In devices where separate transmit and receive antennas are deployed [4]–[6], the antenna separation provides natural isolation typically in the order of 20 dB to 30 dB between the transmit and receive chains. In shared-antenna devices, on the other hand, circulator based approaches [3], [7] as well as structures based on electrically balanced hybrid-junction circuits [8]–[10] have been reported, with isolation numbers ranging typically in the order of 20 dB to 40 dB. Furthermore, the idea of deploying dual-port polarized antenna system is reported in [9].

On top of passive isolation, also active RF cancellation is commonly deployed to further suppress the self-interference prior to entering the sensitive receiver chain [3]–[5], [7]–[9]. Various different concepts and solutions for active RF cancellation have been reported, see, e.g., [3]–[5], [8], [9]. Most of the reported works are, however, limited to fairly narrow bandwidths in the order of 1 MHz to 20 MHz or so, and are commonly also building on the assumptions of highly linear RF components and static self-interference channel whose characteristics do not change over time. In this work, we describe a solution that can provide wideband RF cancellation, operate under highly nonlinear mobile power amplifier, and contains also self-adaptive or self-healing properties to automatically follow and track the changes in the self-interference channel. Such changes can in practice occur due to, e.g., changes in the reflections from the surrounding objects or from the antenna reflections in shared-antenna based devices. Compared to the existing literature, we report measurement results of the developed RF cancellation architecture up to 100 MHz instantaneous waveform bandwidths in both shared-antenna and dual-antenna configurations, and using off-the-shelf RF components, reaching around 20 dB to 40 dB cancellation levels under the self-adaptive control. Furthermore, tracking against time-varying self-interference channel characteristics is demonstrated. No such results have been reported earlier in the existing literature.

The rest of the paper is organized as follows. In Section II, we describe the assumed overall device architecture. Section III, in turn, describes the principal RF cancellation circuit architecture, self-adaptive control and the corresponding mathematical modeling. In Section IV, we describe the demonstration circuit implementation and corresponding RF measurement results. Section V will finally conclude the paper.

This work was funded by the Academy of Finland (under the project #259915 ”In-band Full-Duplex MIMO Transmission: A Breakthrough to High-Speed Low-Latency Mobile Networks”), the Finnish Funding Agency for Technology and Innovation (Tekes, under the project ”Full-Duplex Cognitive Radio”), the Linz Center of Mechatronics (LCM) in the framework of the Austrian COMET-K2 programme, and by the Internet of Things program of DIGILE (Finnish Strategic Centre for Science, Technology and Innovation in the field of ICT) funded by Tekes.

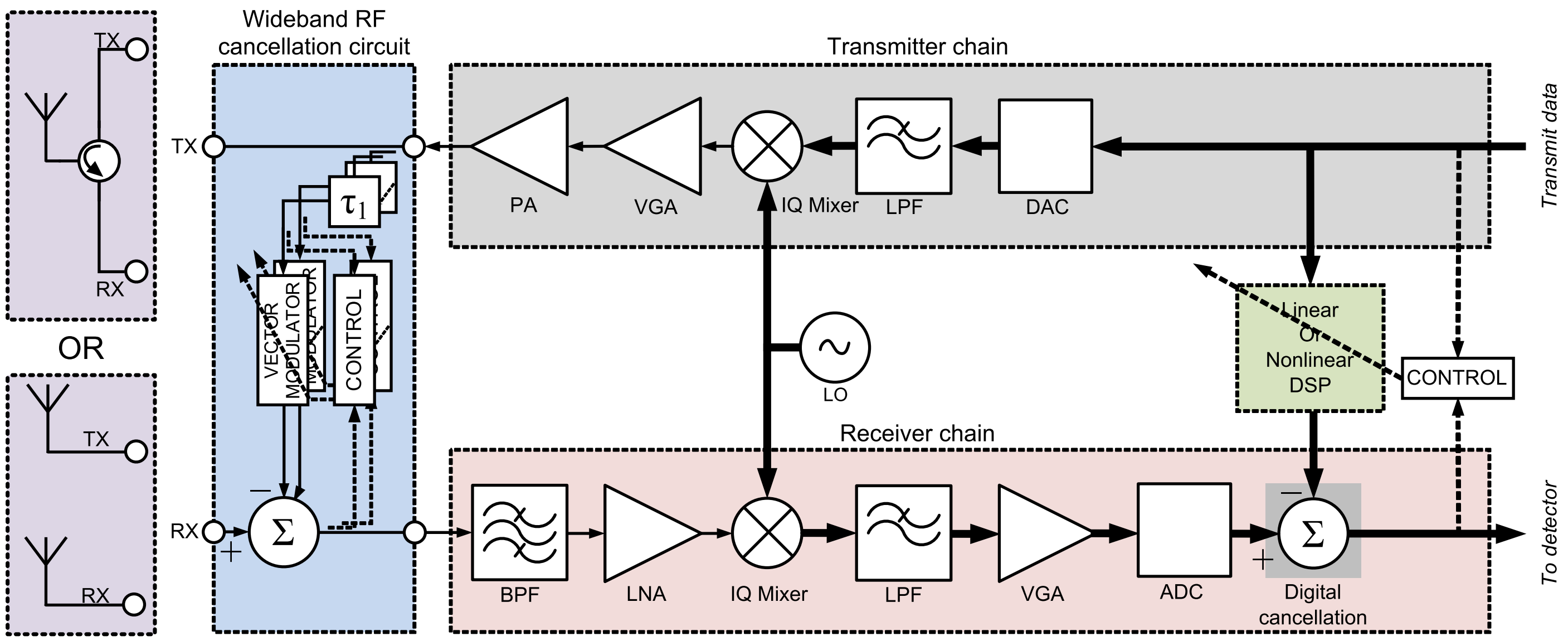

Fig. 1. The considered full-duplex device architecture. On the left-hand side, both shared-antenna and dual-antenna based setups are shown.

## II. Full Duplex Device Architecture

The overall assumed full-duplex device architecture is shown in Fig. 1. In terms of the transceiver chains, it builds on the widely-used direct-conversion radio architecture, which is well suitable especially for compact low-cost radios but also widely applied in high-end products such as base-station devices. As the figure illustrates, there are two options for the actual antenna interface: shared-antenna and dual-antenna configurations. In both cases, active RF cancellation, elaborated in more details in Section III, is deployed, building on injecting a cancellation signal at the input of the sensitive receiver chain to suppress the self-interference. The reference for the cancellation is taken from the power amplifier (PA) output, which offers robustness against transmit chain imperfections [11]–[13]. In addition to RF cancellation, also digital cancellation is in practice needed to further suppress the self-interference below the system noise floor, prior to feeding the received signal towards detector. In the literature [3], [7], [11]–[13], both linear as well as nonlinear digital cancellation processing solutions have been proposed and evaluated, while in this work we fully focus on the advances in the active RF cancellation stage, leaving the analysis of the overall integrated cancellation performance for future work. As shown in [12], high-performance RF cancellation is very important to protect the sensitive receiver circuitry and thus to avoid nonlinear distortion in the receiver.

When considering the self-interference coupling channel between the PA output and low-noise amplifier (LNA) input, the antenna configuration has an essential role. In a shared-antenna based device, the direct leakage through a circulator (or a hybrid junction based circuit, as in [8], [9]) forms one central self-interference source [14]. However, with typical antenna matching levels, the reflection from the antenna can be even more substantial and is attenuated by only 20 dB or so, depending on the antenna matching [15]. This self-interference component has also a longer delay compared to the direct leakage. Finally, there are also additional reflections coming from the surrounding objects and environment, with even longer delays. However, these reflections are also substantially weaker due to higher path losses.

In the case of a dual-antenna device with separate antennas for transmission and reception, the direct coupling between the antennas is naturally the strongest source of self-interference, having also the shortest delay. The attenuation of this component is primarily stemming from the used frequencies and physical antenna separation. On top of this component, there are then again also reflections from the surrounding objects, constituting additional self-interference components with longer delays. Thus, in both shared-antenna and dual-antenna devices, the overall self-interference coupling channel is of multipath nature, while the power delay profiles are different. With wideband waveforms, such multipath channel maps to frequency-selective self-interference coupling, which needs to be taken into account both in the RF cancellation stage as well as in the digital canceller. Furthermore, to support efficient cancellation of the self-interference under practical conditions in time-varying environments, the control of the phases and amplitudes of the RF cancellation signals needs to be self-adaptive in order to track sudden changes in the close proximity of the antenna. In general, by monitoring the signal instantaneous power level at the canceller output, this control can be made automatic by using either digital or analog circuits [7]. Such self-tuning of the analog/RF cancellation signal is perhaps one of the most crucial elements of a feasible in-band full-duplex radio, especially on the mobile device side, but is typically neglected in most of the reported works in this field so far.

## III. RF Cancellation Circuit Architecture

This section provides the basic architecture and operating principle of the wideband self-adaptive RF canceller. For notational convenience, we provide the basic modeling using complex-valued baseband equivalent signals, while the actual cancellation signal processing takes place at RF. Also the actual useful received signal and external interference signals are neglected for notational convenience when modeling the signal in the receiver chain.

### A. Operating Principle and Modeling

We start by writing the self-interference signal prior to RF cancellation as

$$y(t) \ = \ h(t) * x(t) + n(t), \tag{1}$$

where $x(t)$ is denoting the transmitted signal at PA output, $h(t)$ is the overall effective multipath coupling channel, $n(t)$ is noise, and $*$ denotes the convolution operation. Notice that in this notation, $x(t)$ contains all the imperfections of the actual transmit chain. Then, the RF canceller itself builds on a set of fixed delays and adjustable amplitude and phase control per delay branch, which maps to complex coefficients at baseband equivalent modeling level. Hence, we write the RF canceller output signal as

$$z(t) \ = \ y(n) - \sum_{n=1}^{N} w_n x(t - \tau_n), \tag{2}$$

where $\tau_n$ denotes the fixed delay of the $n^{th}$ branch and $w_n$ is a complex-valued coefficient modeling the corresponding amplitude and phase change.

This principal processing is depicted in Fig. 2. (left subfigure) where vector modulators are implementing the amplitude and phase control in the parallel branches. Furthermore, the circuit incorporates an additional LNA prior to cancellation point, to guarantee that there is sufficient power available for the cancellation signal when there are strong reflections, e.g., from the antenna in the shared-antenna device. In general, since the implemented RF canceller is building on a set of fixed delays, this kind of RF cancellation architecture can be interpreted as an RF interpolator, which is seeking to accurately regenerate the exact SI waveform and then subtract it from the receiver observation.

The use of vector modulators for tuning the cancellation signals also allows for a significant reduction in the number of delay lines. In [3], where only the amplitudes of the cancellation signals are tuned with variable attenuators, 16 delay lines are required, as opposed to the two delay lines used in our proposed architecture. This is due to the fact that the vector modulators can also tune the phases of the cancellation signals, which considerably improves the modeling accuracy.

### B. Control and Self-Adaptive Tuning Features

Accurate regeneration and cancellation of the self-interference requires careful optimization and control of the amplitudes and phases of the parallel branches. In this work, automatic self-tuning features are enabled through analog least mean square (LMS) type adaptive filtering algorithm, which is seeking to minimize the instantaneous squared error at the canceller output. For fast speed and tracking capabilities, fully analog implementation is deployed, contrary to classical digital signal processing (DSP) based implementations where also additional data converters would be needed.

For the analog tracking through LMS, the canceller instantaneous output signal $z(t)$, IQ demodulated to baseband, is deployed as the feedback signal. This is denoted below by $z_{IQ}(t)$. Furthermore, a similar IQ demodulated baseband observation of the delayed TX signal, denoted by $x_{IQ}(t-\tau_n)$, is also deployed, yielding together an analog LMS learning rule of the form

$$w_n \leftarrow w_n + \mu \int x^*_{IQ}(t - \tau_n) z_{IQ}(t)\, dt, \tag{3}$$

where $\mu$ denotes the learning step-size controlling the learning and tracking rate as well as the steady-state variance. This can be written in terms of the parallel I and Q signals and quantities as

$$w_{n,I} \leftarrow w_{n,I} + \mu \int \left(x_I(t - \tau_n) z_I(t) + x_Q(t - \tau_n) z_Q(t)\right)\, dt \tag{4}$$

$$w_{n,Q} \leftarrow w_{n,Q} + \mu \int \left(x_I(t - \tau_n) z_Q(t) - x_Q(t - \tau_n) z_I(t)\right)\, dt \tag{5}$$

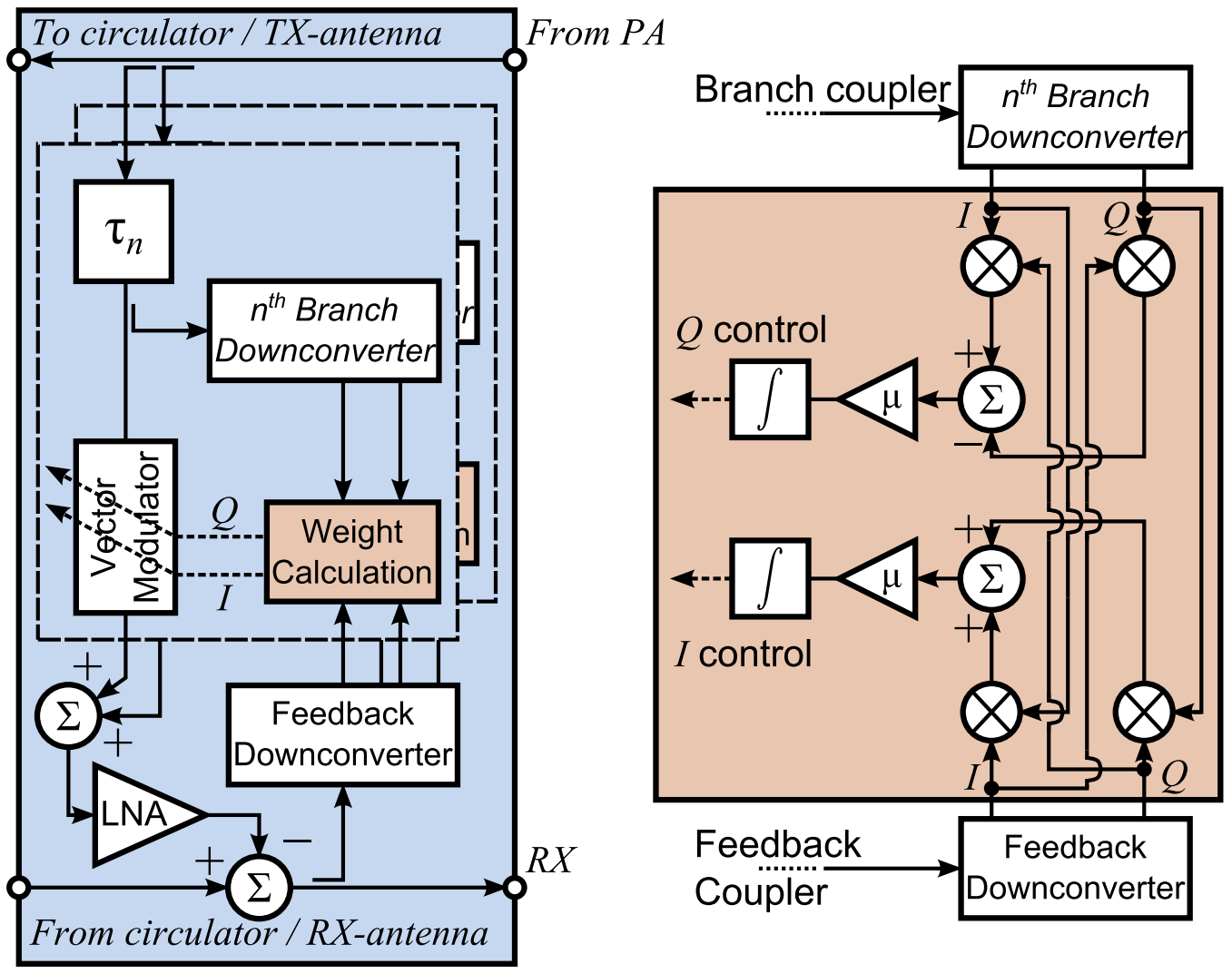

Fig. 2. Multi-tap RF canceller with fixed delays and adjustable amplitude and phase control per branch. Also shown on the right-hand side is the self-adaptive analog control-loop for the I and Q control voltages of an individual vector modulator.

This is illustrated in terms of principal circuit architecture in Fig. 2 (right subfigure).

## IV. Implementation and Measurement Results

### A. Demonstrator Circuit Implementation

The actual demonstration circuit for the developed RF canceller is built following the principal structures in Fig. 2. The blue areas are portions of the circuit operating at RF-frequencies and the light brown colored portion is operating at analog baseband. The demonstration board is built using Arlon 25FR RF printed circuit board (PCB) material using mostly surface mount devices (SMD). Two cancellation branches are implemented in this demonstration canceller, the size of the finished board being 10 cm × 12 cm. In addition, integrators and the LNA are constructed on small separate boards in order to make the overall canceller more versatile.

In general, it should also be emphasized that the aim of this design is only to demonstrate the concept and its operating principle, and to carry out first RF performance measurements. Therefore, many parameters such as TX and RX chain insertion losses and output noise are not yet fully optimized, but will be done in later implementation rounds. Lacking IC prototyping capabilities, the design remains nonetheless complex as only off-the-self components can be used.

The demonstration board is implemented as follows. TX-signal is passed through the board to TX-antenna or circulator and a part of the signal is coupled using an SMD directional coupler (Anaren DC2337J5010AHF). Then the signal is delayed using a shared delay line (Anaren XDL15-2-020S) and divided after that using a Wilkinson power divider (Anaren PD2328J5050S2HF). The second branch has an additional delay implemented using a coaxial cable interconnection, which allows evaluation of different delays for the second canceller branch. Other than that the branches are similar and consist of an analog vector modulator (Hittite HMC631LP3) and a control circuit.

Control voltages are created using the block diagram depicted in Fig. 2. (right subfigure). First, the signals are downconverted to baseband IQ signals (Maxim Integrated MAX2023) and then routed to multiplication integrated circuits (IC) (Analog Devices AD835). These multiplier ICs provide also a summing node and are utilized as described in the block-diagram of Fig. 2. The subtraction is done by simply inverting one of the signals before multiplication.

After the multiplication, the signal is low-pass filtered. This is done using operational amplifiers (Texas Instruments OPA2735) and the well-known inverting integrator circuit. An inverting buffer amplifier before the integrator makes the cascade non-inverting as intended. It is also used to feed a DC offset voltage such that it nulls all the unwanted DC-offsets due to the component impairments. Currently, this adjustment is done manually. The chosen operational amplifier is of chopper type which has a particularly low DC-offset. This is critical as the integration circuit has a DC-gain of 40 dB to 60 dB and thus even the smallest DC error in the input can yield substantial error at the output.

The integration DC-gain and bandwidth can be adjusted on the fly using a potentiometer and a capacitor bank, respectively. Finally, the control voltage is routed to the vector modulators. Note that it is possible to control the vector modulators also manually using potentiometers, which are visible in Fig. 3.

The cancellation signals are then combined and amplified. An LNA (Avago MGA-638P8) is required in order to have enough cancellation power for the strongest SI components and to compensate for the losses of the cancellation taps. When testing with high intrinsic isolation full-duplex setups, e.g. dual-antenna, the LNA can be replaced by a short coaxial cable to lower the output noise. The signal is at this point 180 degrees out of phase and is then combined with the actual received signal by using a Wilkinson power divider. This is essentially a lossy combiner but it allows the combining of signals with arbitrary phases. Before the actual RX-chain, a part of the canceller output signal is taken for feedback using a directional coupler. In fact, the idea of the canceller is to minimize this feedback signal and hence reduce also the SI power. Fig. 3. depicts the built demonstration canceller without the integration circuits.

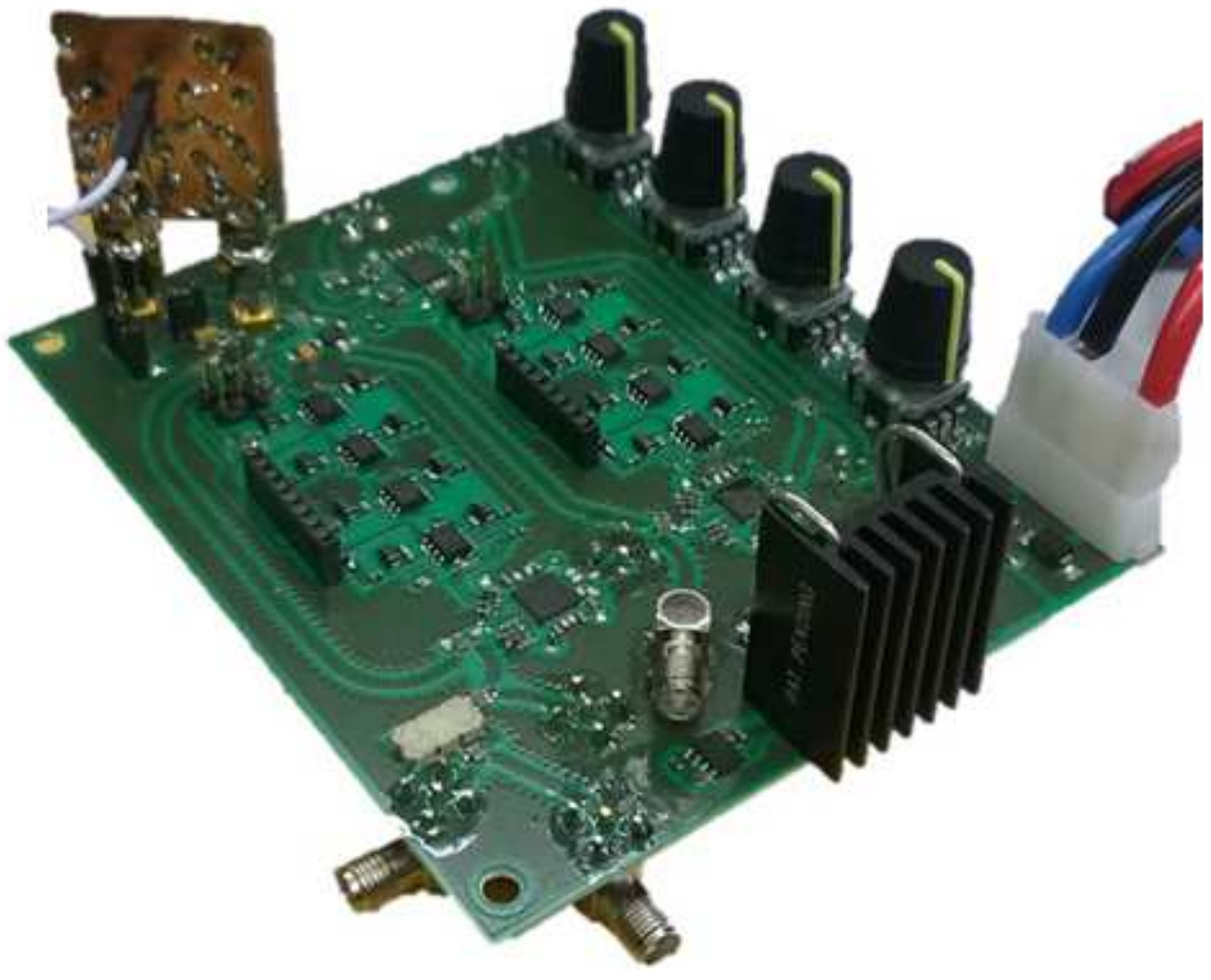

Fig. 3. The RF canceller demonstration board. On top right is the manual control, on top left are the LNA board and the vector modulators, in the middle is the analog self-adaptive control circuit together with the slots for the integrator cards, and at the bottom right is the power supply.

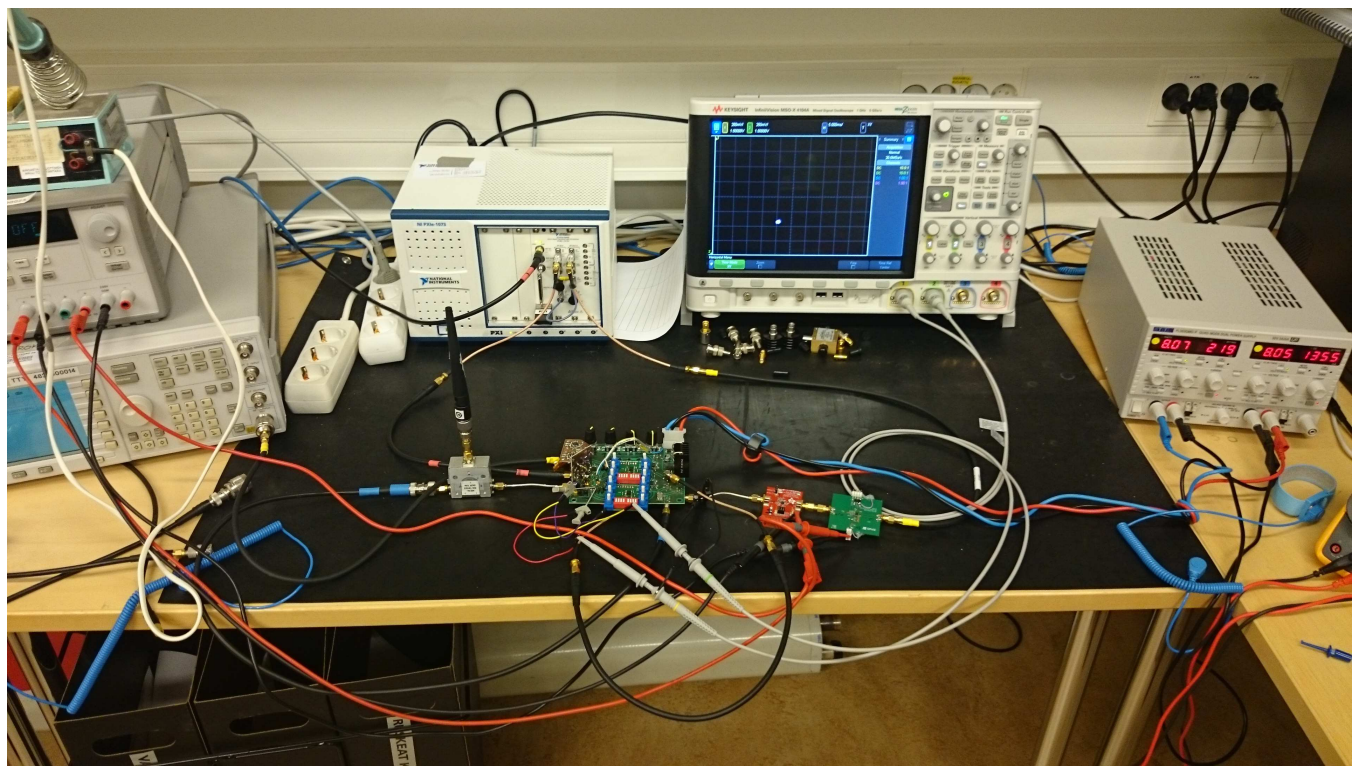

Fig. 4. Overall RF measurement setup.

### B. Measurement Setup

In order to evaluate the performance of the built canceller, a measurement test setup was built in standard laboratory conditions. This makes the testing more realistic, compared to, e.g., an anechoic chamber, as there will be also multiple reflections from nearby objects. Fig. 4. depicts the overall measurement setup in the circulator-based shared antenna configuration. In all the measurements, the operating frequencies are at the 2.4 GHz ISM band.

National Instruments PXIe-5645r is used as the signal generator and receiver. The TX signal is first generated at low power and then boosted using RFMD RF5125 and finally amplified to +21 dBm using TI CC2595. This signal is then fed to the canceller and is heavily distorted due to the low-cost PA driven already substantially beyond the 1 dB compression point. After the RF-cancellation, the output spectrum and power are monitored on the computer screen. LO-source for the canceller is HP E4437B ESG-DP and IQ drive signal is monitored using Keysight MSO-X 4104A. In the shared-antenna measurements, a MESL SG 10522 circulator is used with a Cisco AIR-

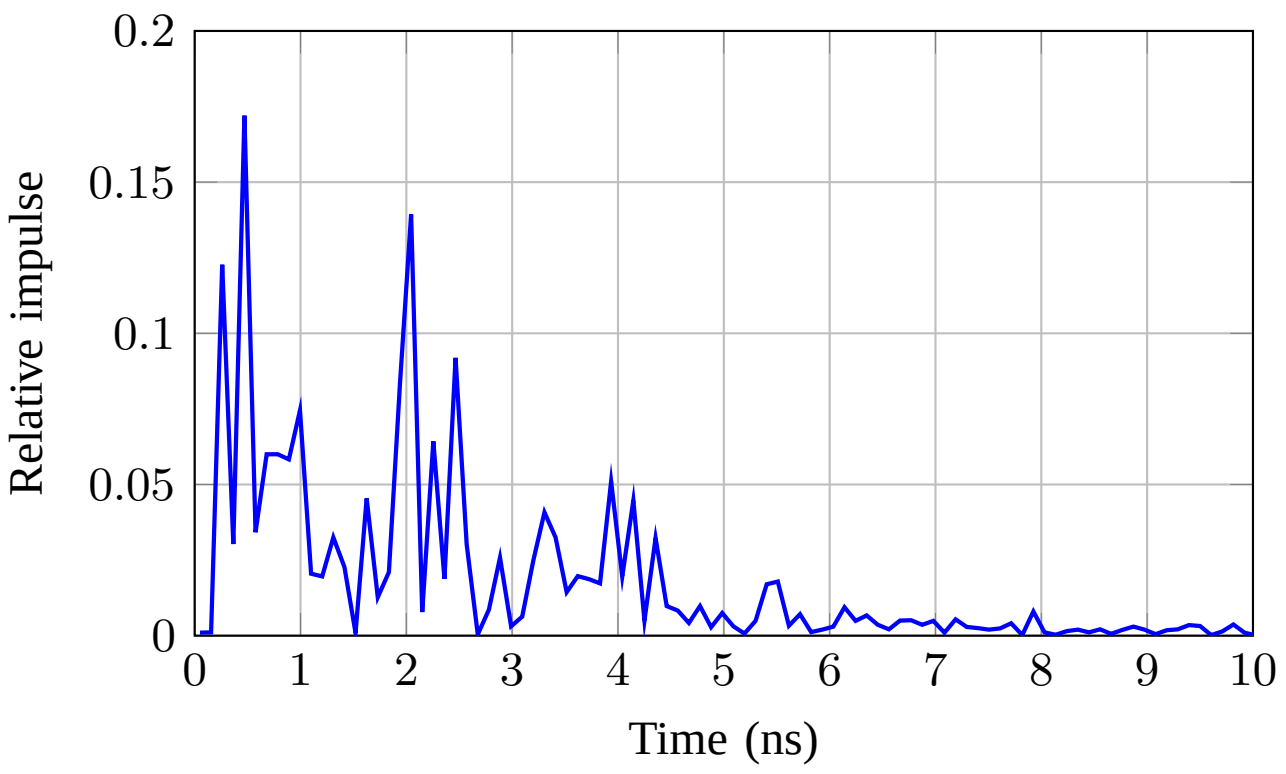

Fig. 5. Measured impulse response of the used circulator with antenna connected without cabling. Note that this measurement method, VNA based impulse response, provides only accurate time data whereas the absolute values of the amplitudes are only indicative.

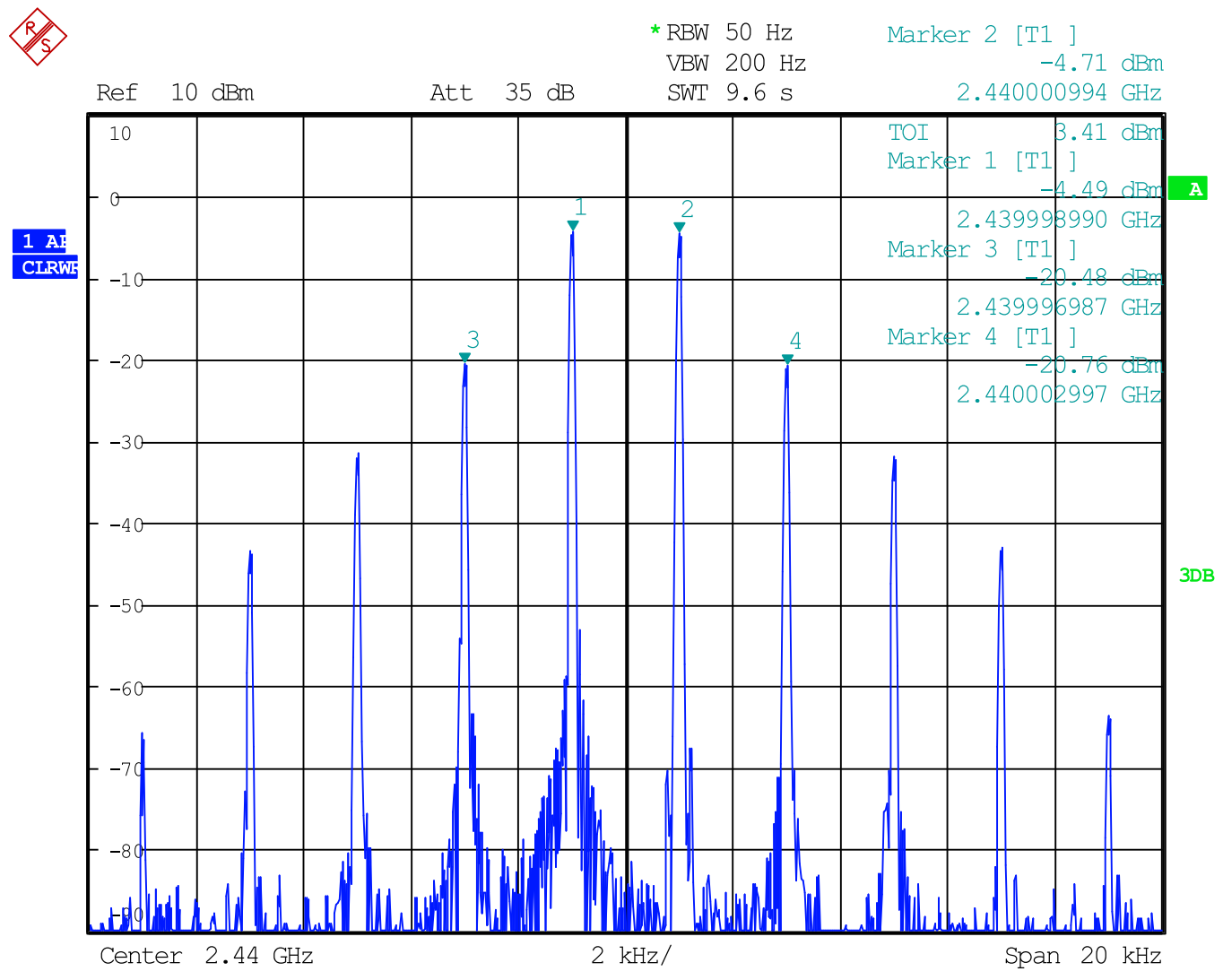

Fig. 6. Measured two-tone test output signal of the used PA with 0 dBm input power, corresponding to 21 dBm output power. Values shown here are measured through a 20 dB attenuator in order to protect the spectrum analyzer.

ANT494i antenna. For the dual antenna setup, two Cisco antennas are used with a separation of 30 cm. The antennas are positioned in a co-polarized way.

To illustrate some essential characteristics of the shared-antenna device, an effective measured impulse response of the circulator setup, with the antenna connected, is shown in Fig. 5. As can be seen, the delay of the direct coupling through the circulator is around 0.5 ns, while the strong antenna reflection has a delay of approximately 2 ns. Taking into account the actual interconnections on the demo board, the fixed delays in the two-branch RF canceller are set to 5 ns and 7.5 ns as the most significant self-interference components are within this delay spread. This allows for accurate regeneration of the total self-interference signal in the RF canceller [7]. Of these values, already 3 ns comes from the intrinsic delay of the components and transmission lines. The same setup is also used in the dual-antenna measurements where the delay of the direct connection between the two antennas, with cabling, fits to this interval. Furthermore, Fig. 6 shows an example two-tone test measurement of the used mobile PA, demonstrating that it is operating in a heavily nonlinear region.

### *C. Circulator-Based Measurement Results*

The cancellation performance with the circulator setup was measured using band limited signals with both 20 MHz and 100 MHz instantaneous bandwidths at the 2.4 GHz ISM band. The RF canceller was configured such that branch one was in self-adaptive control and branch two in manual mode in order to find the best performance.

The cancellation performance with a 20 MHz bandwidth is depicted in Fig. 7(a) Alongside with the cancellation results, the original TX signal and the self-interference signal after the circulator (intrinsic attenuation) are shown. Note the high distortion after the low-cost mobile PA, which was already evident in Fig. 6. There is also a rather strong LO leakage present, which is due to the used NI transceiver. The canceller is able to reduce the SI power down to −38 dBm over the 20 MHz bandwidth, thus providing a total of 59 dBs of cancellation. Out of this, the actual active RF cancellation contributes 33 dB. Fig. 7(b) then presents the corresponding results for the 100 MHz bandwidth case. Total cancellation performance is 41.5 dB, consisting of 23.5 dB of intrinsic attenuation and 18 dB of active RF-cancellation. The results are promising, given that the canceller implementation is not yet optimized for bandwidths over 20 MHz.

The increase in the power outside the nominal carrier bandwidth in Fig. 7(a) is due to the applied cancellation signal. The distorted PA output is fed to the vector modulator which has now been set to optimize cancellation for the gray shaded area. Hence, the power in the edges of the output spectrum after the cancellation may increase due to residual PA distortion stemming from the cancellation network. More branches would increase the cancellation for wider bandwidths.

As a reference, in [3], deploying a total of 16 canceller branches with computer-based digital amplitude control per branch, and relying also on known WiFi-type preamble signals to learn the coupling channel offline, a total of 70 dB cancellation was obtained over a bandwidth of 20 MHz. The design proposed in this paper achieves good performance by using only two delay lines, as it is capable of tuning also the phases of the reference signals, unlike the canceller in [3], which only tunes the amplitudes. This allows for a significant reduction in the number of delay lines, while still providing good performance, as already discussed. The potential of our design is thus very promising, building on only two parallel branches instead of 16 and using self-adaptive online analog control without any known preambles or calibration periods, while also operating under highly nonlinear low-cost mobile PA. Furthermore, the automated tracking capabilities are demonstrated in Subsection IV-F.

### *D. Dual Antenna Based Measurement Results*

The performance was tested similarly with dual antennas. A significant difference in this case is that, instead of two strong components in the SI delay spread, now only one dominant component is present due to the direct line of sight coupling between the TX and RX antennas. Fig. 8(a) presents the results for the 20 MHz bandwidth case. Here, the intrinsic cancellation is stronger than in the circulator setup due to the path loss between the antennas. The canceller works equally well for the dual antenna case, reducing the SI power down to −43.6 dBm over a 20 MHz bandwidth. In total, the cancellation is close to 65 dB, of which roughly 28 dB is contributed by the active RF cancellation. The improvement in performance with respect to the shared-antenna case can most likely be contributed to the different self-interference signal structure, as there is now only one significant self-interference component instead of two. This obviously makes it easier to produce an accurate cancellation signal.

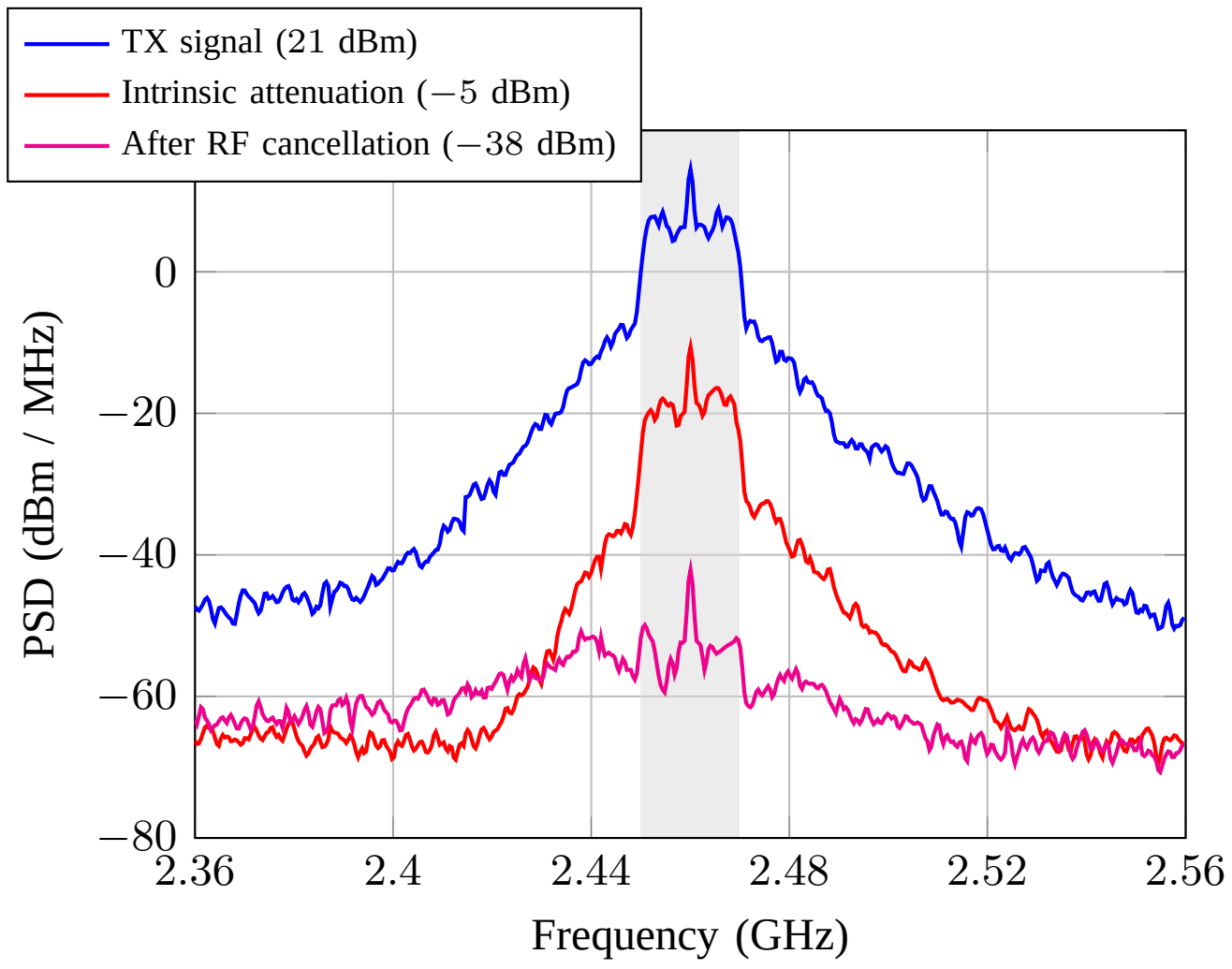

(a)

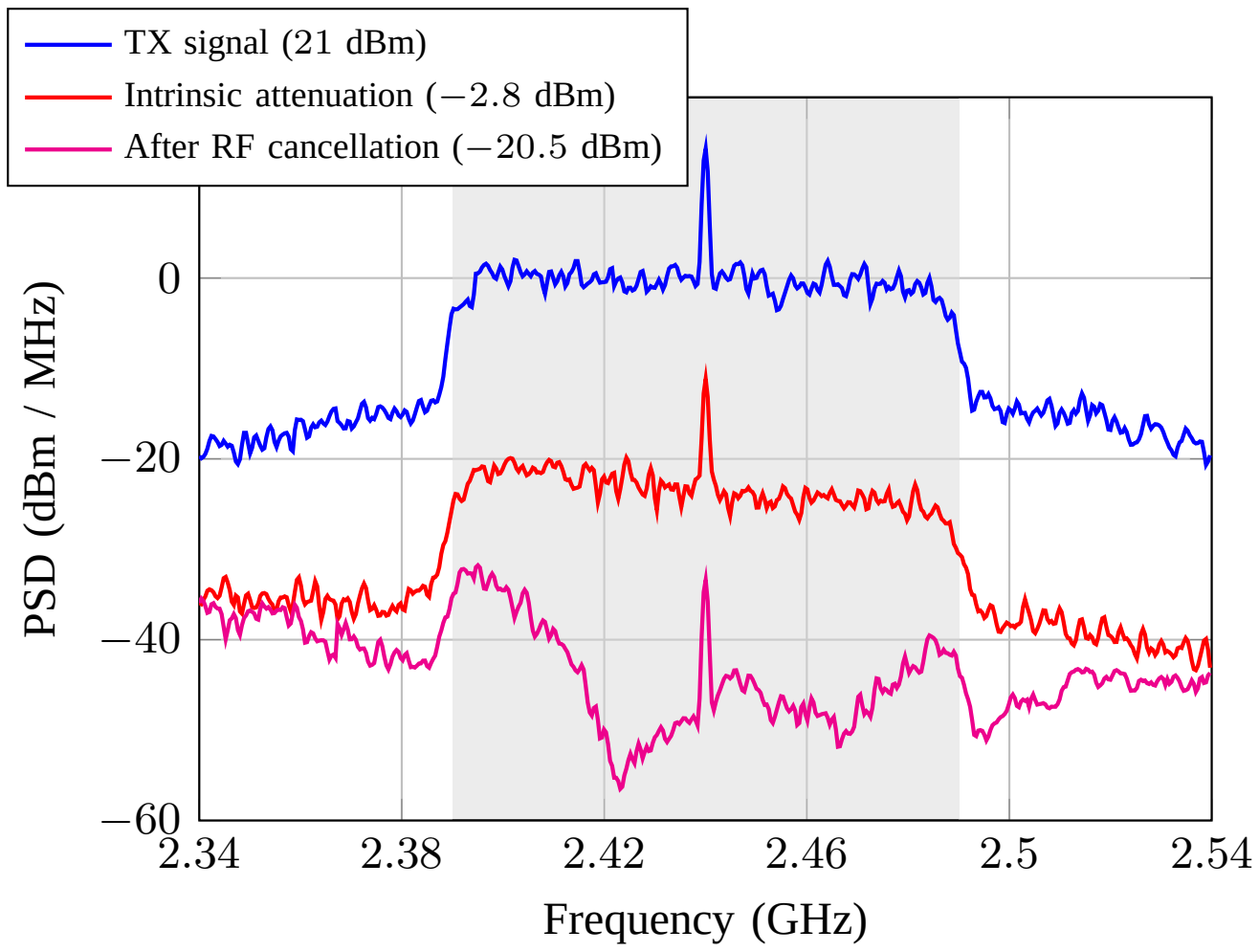

(b)

Fig. 7. Measured RF cancellation performance with circulator-based shared-antenna device using (a) 20 MHz and (b) 100 MHz instantaneous bandwidth at 2.4 GHz ISM band.

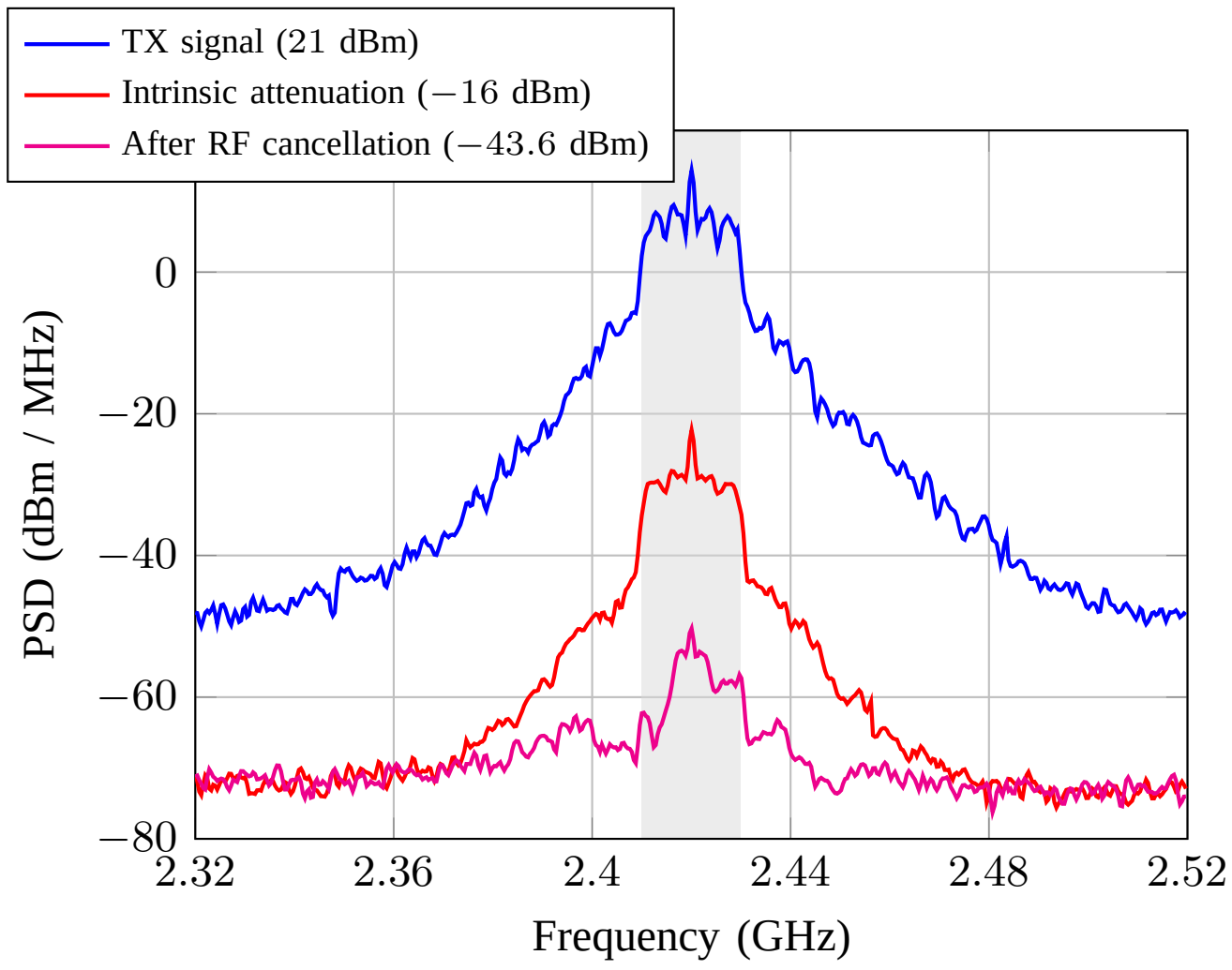

(a)

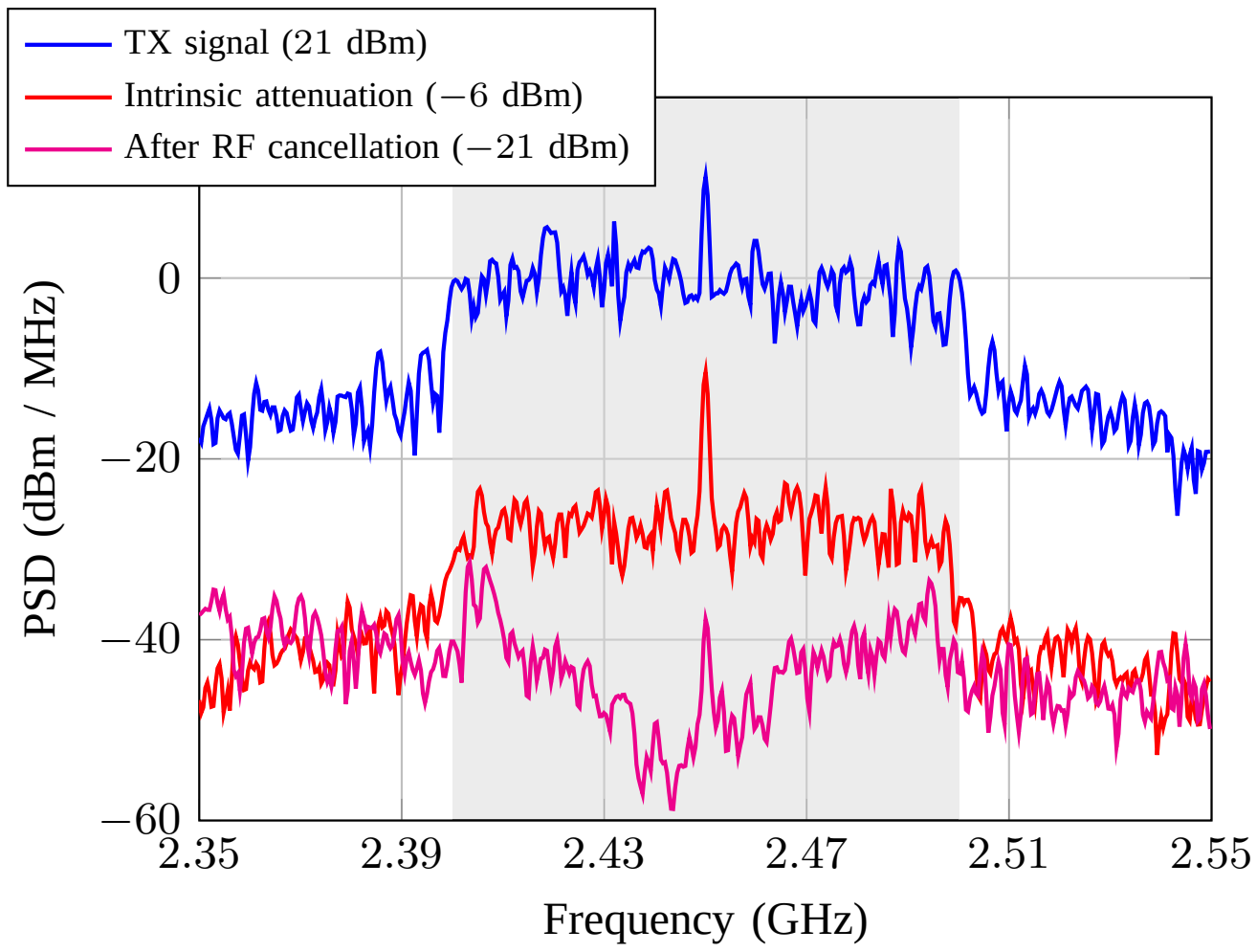

(b)

Fig. 8. Measured RF cancellation performance with dual-antenna based device using (a) 20 MHz and (b) 100 MHz instantaneous bandwidths at 2.4 GHz ISM band.

Figure 8(b) presents the corresponding results obtained with a 100 MHz waveform bandwidth. Here the total cancellation is 42 dB, half of which is intrinsic attenuation and half is active RF cancellation. Thus, with the wider bandwidth, the performance of the dual antenna based system is approximately equal to the circulator-based setup, most likely due to the fact that now the bandwidth is the limiting factor, instead of the self-interference signal structure.

### *E. Recovering Signal of Interest*

To demonstrate that the canceller does not alter the signal of interest (SoI), and can in fact recover it properly, Fig. 9 presents a measurement example with the SoI also present, using the circulator-based setup. As can be seen, the SOI is not affected by the canceller, and, correspondingly, the performance of the canceller is approximately similar as without the SOI. Note that, due to the narrow bandwidth of the SOI in this experiment, the resolution bandwidth in Fig. 9 is 100 times smaller than in the other figures. A video was also recorded from these measurements to further demonstrate the effect of the SOI, please see Subsection IV-F.

### *F. Tracking capabilities*

To demonstrate the tracking capabilities, the canceller is run in fully self-adaptive mode while deliberately disturbing the surroundings of the antenna. The canceller is able to quickly follow the changes and drive the corresponding IQ control voltages such that that a high level of cancellation is maintained. The following website contains videos demonstrating this with single and dual branch adaptive cancellation examples.

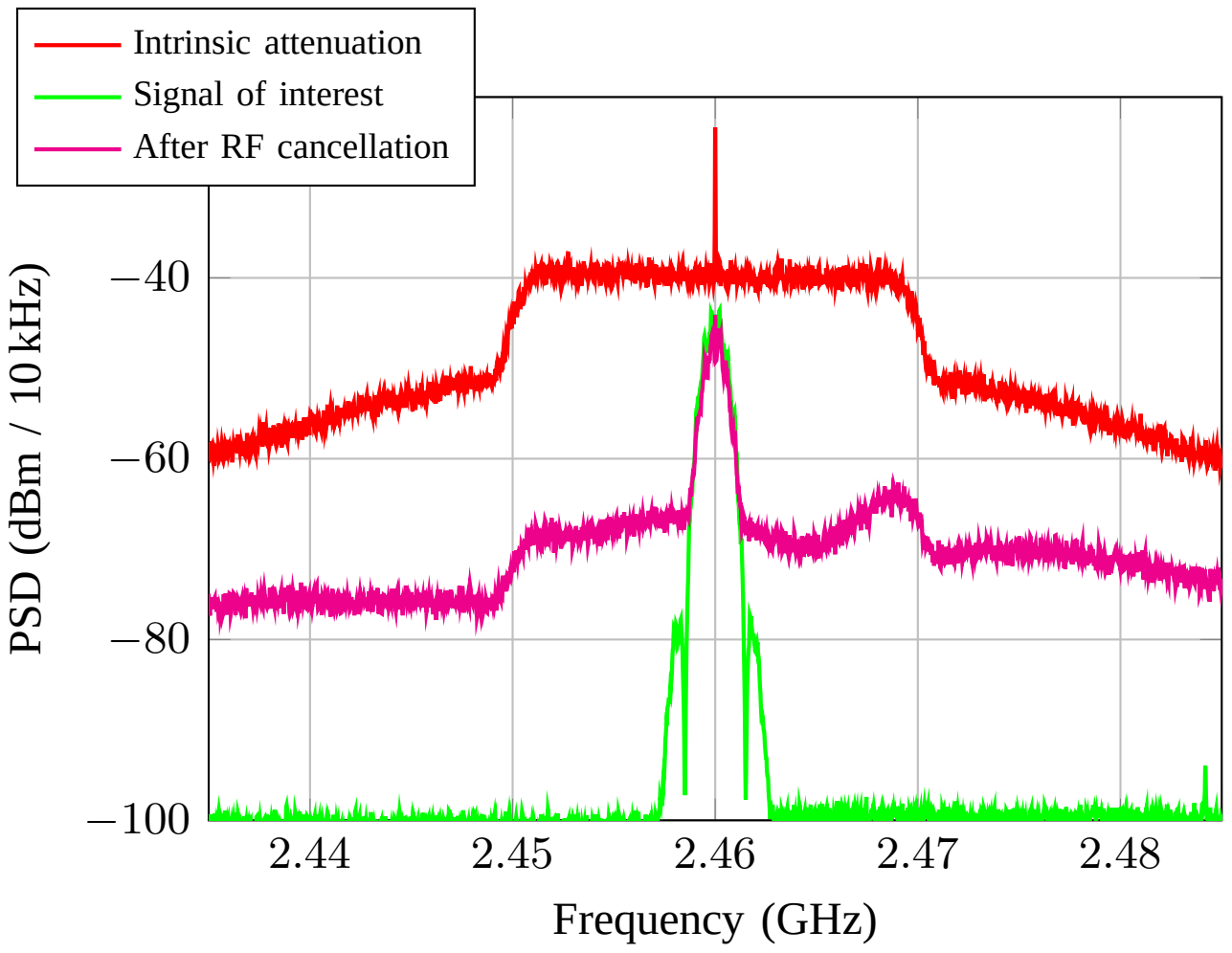

Fig. 9. Measured spectra including also signal of interest (SoI), originally buried under self-interference. The bandwidth of the SoI is deliberately set more narrow than the self-interference signal, in order to enhance the visual insight in the experiment.

- http://www.tut.fi/full-duplex/videos

## V. Conclusion

This paper addressed recent advances in active self-interference cancellation at RF in full-duplex radio transceivers. A novel multi-branch RF canceller circuit architecture was described, suitable for wideband cancellation, and containing self-adaptive or self-tracking automatic control for maintaining high cancellation levels in time-varying reflection scenarios. Complete demonstration implementation and corresponding RF measurements were also described, evidencing successful cancellation at 2.4 GHz band with both 20 MHz and 100 MHz instantaneous bandwidths, when operating under highly nonlinear low-cost power amplifier. Furthermore, successful online self-adaptive tracking against fast changes in the reflections was demonstrated. Our future work will focus on further enhancing the canceller performance with 100 MHz and beyond waveform bandwidths, as well as integrating also efficient linear and nonlinear digital baseband cancellation algorithms into the overall demonstration setup.